\documentclass[a4paper,aps,superscriptaddress,reprint,showpacs]{revtex4-1}
\usepackage{amsmath, amssymb}
\usepackage{bm, braket}
\usepackage{graphicx}
\usepackage[dvipdfmx,colorlinks=true,linkcolor=blue,citecolor=blue,urlcolor=blue]{hyperref}

\begin{document}
\title{Self-consistent phonon calculations of lattice dynamical properties in cubic SrTiO$_{3}$ with first-principles anharmonic force constants}

\author{Terumasa Tadano}
\email{tadano@solis.t.u-tokyo.ac.jp}
\affiliation{Department of Applied Physics, The University of Tokyo, Tokyo 113-8656, Japan}

\author{Shinji Tsuneyuki}
\affiliation{Department of Physics, The University of Tokyo, Tokyo 113-0033, Japan}
\affiliation{Institute for Solid State Physics, The University of Tokyo, Kashiwa 277-8581, Japan}

\begin{abstract}
  We present an \textit{ab initio} framework to calculate anharmonic phonon frequency and  phonon
  lifetime that is applicable to severely anharmonic systems. We employ self-consistent
  phonon (SCPH) theory with microscopic anharmonic force constants, which are extracted from
  density-functional calculations using the least absolute shrinkage and selection operator
technique. We apply the method to the high-temperature phase of SrTiO$_{3}$ and obtain
  well-defined phonon quasiparticles that are free from imaginary frequencies. 
  Here we show that the anharmonic phonon frequency of the antiferrodistortive mode depends significantly on the system size near the critical temperature of the cubic-to-tetragonal phase transition.
  By applying perturbation theory to the SCPH result, phonon lifetimes are calculated for cubic SrTiO$_{3}$, which are then employed to predict lattice thermal conductivity using the Boltzmann transport equation within the relaxation-time approximation.
  The presented methodology is efficient and accurate, paving the way toward a reliable description of thermodynamic, dynamic, and transport properties of systems with severe anharmonicity, including thermoelectric, ferroelectric, and superconducting materials.
\end{abstract}
\maketitle


\section{Introduction}

Lattice anharmonicity plays an important role in characterizing various physical properties of solids and molecules, 
including the temperature-dependence of vibrational frequencies, thermal expansion and phase stability of solids~\cite{wallace_book}.
It is also responsible for the finite phonon linewidth and the lattice thermal conductivity $\kappa_{\mathrm{L}}$,
which is a key quantity when optimizing the thermoelectric figure-of-merit $ZT$ \cite{2008NatMa...7..105S}.
The magnitude of anharmonicity varies significantly for different materials.
For example, covalent materials such as silicon, diamond and graphene are very harmonic and 
show high thermal conductivities~\cite{PhysRevB.80.125203,PhysRevB.82.115427}.
Conversely, thermoelectric and ferroelectric (FE) materials often show severe anharmonicity, demonstrated by 
inelastic neutron scattering spectra and ultralow $\kappa_{L}$ values~\cite{2011NatMa..10..614D,2014Natur.508..373Z,Takabatake2014}.
Anharmonic effects can also be significant in superconductors~\cite{doi:10.1143/JPSJ.81.011012,Mankowsky:2014em,PhysRevLett.114.157004} 
and materials under extreme conditions~\cite{Luo:2010jg,PhysRevB.89.094109}.
To develop a robust understanding of anharmonic properties of solids, a reliable and versatile computational method is required.
Therefore, the development of first-principles methods to calculate anharmonic properties of 
solids and molecules has been the subject of intense research in recent years.

Many-body perturbation theory is one approach for treating lattice anharmonicity.
This technique considers the anharmonic effects as self-energies~\cite{PhysRev.128.2589}.
The self-energies can be calculated using a systematic approximation to the Feynman diagrams,
where the lowest-order approximation is usually employed in the \textit{ab initio} calculations 
based on density-functional theory (DFT).
Performing this calculation requires the cubic and quartic force constants, 
which are the third- and fourth-order derivatives of the Born-Oppenheimer potential energy surface, respectively.
The third-order terms can be obtained efficiently and systematically using either density functional perturbation theory (DFPT)~\cite{RevModPhys.73.515} or the finite-displacement approach~\cite{Esfarjani2008}.
Using the cubic terms, phonon linewidth can be obtained by evaluating the bubble diagram [Fig.~\ref{fig:selfenergy}(b)].
This type of calculation has been performed to predict the lattice thermal conductivity of many solids~\cite{PhysRevB.80.125203,PhysRevB.82.115427,Tian2012,PhysRevB.91.094306}
and can also be applied to complex materials~\cite{PhysRevLett.114.095501}.
To estimate the phonon frequency shift due to lattice anharmonicity, one also needs to compute the loop diagram [Fig.~\ref{fig:selfenergy}(a)] 
using the quartic terms.
The calculation of the quartic terms can, in principle, be achieved using the finite-displacement approach.
However, since the number of quartic parameters increases rapidly as the number of atoms in the supercell increases, 
such calculations have only been reported for simple systems~\cite{PhysRevB.59.6182,PhysRevLett.99.176802}.

The perturbative approach is valid only when the anharmonic self-energies are sufficiently 
small compared with the harmonic frequency. 
Therefore, one cannot expect this technique to yield accurate results for severely anharmonic systems.
High-temperature phases of FE material are typical cases where the perturbation approach fails because of the
imaginary frequencies of harmonic phonons.
To overcome this limitation, it is necessary to employ a non-perturbative approach to treat anharmonic effects.

Methods based on \textit{ab initio} molecular dynamics (AIMD) can consider anharmonic effects non-perturbatively.
From the velocity-velocity autocorrelation function calculated using the trajectory of an AIMD simulation, 
one can obtain the vibrational density of states with full anharmonicity. 
To obtain the anharmonic frequency and linewidth of individual phonons, 
the velocity should be projected onto the phonon eigenvector~\cite{PhysRevB.89.094109}.
Inherent in this procedure is the assumption that the phonon eigenvectors are not altered by anharmonic effects.
Such an assumption, however, is valid only for simple systems containing a few atoms in the primitive cell.
The temperature-dependent effective potential (TDEP) method \cite{PhysRevB.87.104111} is another AIMD-based approach.
The TDEP method optimizes the \textit{effective} harmonic force constants within an AIMD simulation at a target temperature.
This method should be useful in high temperature because it allows both the phonon eigenvectors and 
the internal coordinate system to be changed by anharmonic effects.
However, since the AIMD is based on the Newton equation of motion, 
the MD-based methods cannot account for the zero-point vibration.
Therefore, these methods cannot be applied to superconductors and ferroelectric 
materials in the low-temperature range.

Self-consistent phonon (SCPH) theory \cite{PhysRevB.1.572} is another approach for including anharmonic effects beyond perturbation theory that considers the quantum effect of phonons.
Other first-principles methods are able to compute anharmonic phonon frequencies related to the SCPH theory: 
self-consistent \textit{ab initio} lattice dynamics (SCAILD) \cite{PhysRevLett.100.095901} and 
stochastic self-consistent harmonic approximation (SSCHA)~\cite{PhysRevB.89.064302}.
To avoid the cumbersome calculation of quartic force constants, these methods employ real-space stochastic approaches and 
displace atoms in the supercell to model anharmonic effects.

In this study, we have developed an efficient first-principles method to treat lattice anharmonicity.
The method is based on the SCPH theory, and the anharmonic frequency is estimated from the pole of the Green's function.
The cubic and quartic force constants necessary for the present SCPH calculations are extracted 
from the DFT calculations using the recently proposed compressive sensing approach~\cite{PhysRevLett.113.185501}.
By combining the perturbation theory with the solution to the SCPH equation, we can also estimate the phonon lifetime and lattice thermal conductivity of severely anharmonic materials.

To confirm the validity of our approach, the method is applied to the high-temperature phase of SrTiO$_{3}$ with cubic symmetry (c-STO). 
SrTiO$_{3}$ is one of the most studied perovskite oxides and is known to undergo the cubic-to-tetragonal phase transition at 105 K 
accompanied by the \textit{freezing-out} of the antiferrodistortive (AFD) soft mode~\cite{PhysRev.177.858,doi:10.1143/JPSJ.26.396,PhysRev.134.A981}.
The FE phase transition is not observed, even at 0 K, because of the zero-point vibration.
Our approach can describe the temperature dependence of the soft-mode frequencies and 
lattice thermal conductivity of the severely anharmonic c-STO.

This paper is organized as follows. First, we introduce the SCPH theory and details of our implementation in Sec.~\ref{sec:scph}. 
We describe the details of the computational conditions, including the compressive sensing of force constants in Sec.~\ref{sec:simulation}.
The method is applied to c-STO and the results are presented in Sec.~\ref{sec:results}.
In Sec.~\ref{sec:scph_STO}, we examine the size- and temperature-dependence of anharmonic phonon frequencies and compare these results with experimental values.
We also calculate the lattice thermal conductivity of c-STO in Sec.~\ref{sec:thermal_conductivity} to show the validity of our approach.
Finally, we conclude this work in Sec.~\ref{sec:conclusion}.

\section{Self-consistent phonon theory}
\label{sec:scph}

\subsection{Potential energy expansion}

The dynamics of interacting ions within the Born-Oppenheimer approximation are described by the Hamiltonian $H=T+U$, 
where $T$ is the kinetic energy and $U$ is the potential energy of the system. 
When $U$ is an analytic function of atomic displacements from equilibrium positions $\{u\}$ , it can be expanded as a Taylor series with respect to $u$ as
\begin{align}
U &= U_{0} + U_{2} + U_{3} + U_{4} + \cdots ,  \label{eq:U_Taylor} \\
U_{n} &= \frac{1}{n!}\sum_{\{\ell,\kappa,\mu\}}\Phi_{\mu_{1}\dots\mu_{n}}(\ell_{1}\kappa_{1};\dots;\ell_{n}\kappa_{n}) \notag \\
& \hspace{30mm} \times u_{\mu_{1}}(\ell_{1}\kappa_{1})\cdots u_{\mu_{n}}(\ell_{n}\kappa_{n}).
\label{eq:Un_IFC}
\end{align}
Here, $U_{n}$ is the $n$th-order contribution to the potential energy, $u_{\mu}(\ell\kappa)$ is the atomic displacement of the atom $\kappa$ in the $\ell$th cell along the $\mu$ direction, and $\Phi_{\mu_{1}\dots\mu_{n}}(\ell_{1}\kappa_{1};\dots;\ell_{n}\kappa_{n})$ is the $n$th-order interatomic force constant (IFC).
In Eq.~(\ref{eq:U_Taylor}) the linear term $U_{1}$ is omitted because atomic forces are zero in equilibrium.

In the harmonic approximation, only the quadratic term $U_{2}$ is considered and cubic, quartic, and higher-order terms are neglected. 
This allows the Hamiltonian $H_{0} = T + U_{2}$ to be represented in terms of the harmonic phonon frequency $\omega$.
To compute the phonon frequency $\omega$, one needs to construct the dynamical matrix
\begin{equation}
D_{\mu\nu}(\kappa\kappa';\bm{q}) = \frac{1}{\sqrt{M_{\kappa}M_{\kappa'}}}\sum_{\ell'}\Phi_{\mu\nu}(\ell\kappa;\ell'\kappa')e^{i\bm{q}\cdot\bm{r}(\ell')},
\label{eq:Dymat}
\end{equation}
where $M_{\kappa}$ is the mass of atom $\kappa$, $\Phi_{\mu\nu}(\ell\kappa;\ell'\kappa')$ are the harmonic IFCs, and $\bm{r}(\ell)$ is a translation vector of the primitive lattice. 
By diagonalizing the dynamical matrix, one obtains harmonic phonon frequencies as
\begin{equation}
\bm{D}(\bm{q})\bm{e}_{\bm{q}j} = \omega_{\bm{q}j}^{2}\bm{e}_{\bm{q}j},
\end{equation}
where the index $j$ labels the phonon modes for each crystal momentum vector $\bm{q}$ and $\bm{e}_{\bm{q}j}$ is the polarization vector of the phonon mode $\bm{q}j$.

\subsection{Dyson equation}

To derive the SCPH equation, we employ the many-body Green's function theory.
The one-phonon imaginary-time Green's function is given as
\begin{align}
G_{\bm{q}j,\bm{q}j'}(\tau)&=\Braket{T_{\tau} A_{\bm{q}j}(\tau)A^{\dagger}_{\bm{q}j'}(0)}_{H} \notag \\
& = Z^{-1}\mathrm{Tr}\{e^{-\beta H} T_{\tau} [A_{\bm{q}j}(\tau)A^{\dagger}_{\bm{q}j'}(0)]\},
\end{align}
where $T_{\tau}$ is the time-ordering operator, $A_{\bm{q}j}(\tau) = e^{\tau H/\hbar} A_{\bm{q}j} e^{-\tau H/\hbar}$ is the displacement operator in the Heisenberg picture, $Z = \mathrm{Tr} e^{-\beta H}$ is the partition function, and $\beta = 1/kT$, where $k$ is the Boltzmann constant and $T$ is the temperature. The displacement operator is defined as $A_{\bm{q}j} = b_{\bm{q}j} + b_{-\bm{q}j}^{\dagger}$ where $b_{\bm{q}j}$ and $b_{\bm{q}j}^{\dagger}$ are the annihilation and creation operators of the phonon $\bm{q}j$, respectively. It is straightforward to show that the Green's function satisfies $G_{\bm{q}jj'}(\tau) = G_{\bm{q}jj'}(\tau + \beta\hbar)$ for $-\beta\hbar < \tau < 0$ and $G_{\bm{q}jj'}(\tau) = G_{\bm{q}jj'}(\tau-\beta\hbar)$ for $0 < \tau < \beta\hbar$, where we simply denote $G_{\bm{q}j,\bm{q}j'}$ as $G_{\bm{q}jj'}$. Because of these properties, we can also show the following result for the Fourier transform of the Matsubara Green's function:
\begin{equation}
G_{\bm{q}jj'}(i\omega_{m}) = \int_{0}^{\beta\hbar} d\tau G_{\bm{q}jj'}(\tau)e^{i\omega_{m}\tau},
\end{equation}
where $\omega_{m} = 2\pi m/\beta\hbar$ is the Matsubara frequency. 
To obtain the Green's function for anharmonic systems, we need to solve the Dyson equation. 
When one obtains $G_{qjj'}(i\omega_{m})$ within some approximations, it is possible to obtain the retarded Green's function $G_{\bm{q}jj'}(\omega)$ by analytic continuation to the real axis as $G_{\bm{q}jj'}(\omega) = G_{\bm{q}jj'}(i\omega_{m}\rightarrow \omega + i\epsilon)$ with a positive infinitesimal $\epsilon$. 
The function $G_{\bm{q}jj'}$ has a pole at the energy corresponding to the renormalized frequency $\Omega_{\bm{q}j}$.
In the case of the harmonic approximation, one can readily obtain the expression for $G_{\bm{q}jj'}(\omega)$ as
\begin{equation}
G^{0}_{\bm{q}jj'}(\omega) = -\frac{2\omega_{\bm{q}j}}{\omega^{2}-\omega_{\bm{q}j}^{2}}\delta_{jj'}.
\end{equation}
Therefore, the free-phonon Green's function is diagonal in the phonon polarization index $j$ and can be obtained from the harmonic phonon frequencies.

To estimate the phonon Green's function $G_{\bm{q}jj'}(\omega)$, and thereby obtain the anharmonic frequency $\Omega_{\bm{q}j}$, we solve the Dyson equation 
\begin{equation}
[\bm{G}_{\bm{q}}(\omega)]^{-1} = [\bm{G}^{0}_{\bm{q}}(\omega)]^{-1} - \bm{\Sigma}_{\bm{q}}(\omega).
\label{eq:Dyson}
\end{equation}
Here we denote the retarded Green's functions in the matrix form and $\bm{\Sigma}_{\bm{q}}(\omega)$ is the phonon self-energy, which can be estimated within a systematic diagrammatic approximation. 
Since the left-hand side of Eq.~(\ref{eq:Dyson}) becomes zero at the frequencies of the renormalized phonons,
finding the solution $\{\Omega_{\bm{q}j}\}$ is equivalent to solving the following equation
\begin{equation}
\det{\{[\bm{G}^{0}_{\bm{q}}(\omega)]^{-1} - \bm{\Sigma}_{\bm{q}}(\omega)\}} = 0.
\label{eq:Secular}
\end{equation} 
By multiplying $\det{(\bm{\Lambda}_{\bm{q}}^{\frac{1}{2}})}$ from the left and right of Eq.~(\ref{eq:Secular}) with the diagonal matrix $\Lambda_{\bm{q}jj'} = 2\omega_{\bm{q}j}\delta_{jj'}$, one obtains the following SCPH equation:
\begin{align}
&\det{\{\omega^{2} - \bm{V}_{\bm{q}}(\omega)\}} = 0, \\
&V_{\bm{q}jj'}(\omega) = \omega_{\bm{q}j}^{2}\delta_{jj'} - (2\omega_{\bm{q}j})^{\frac{1}{2}}(2\omega_{\bm{q}j'})^{\frac{1}{2}}\Sigma_{\bm{q}jj'}(\omega). \label{eq:scph}
\end{align}
This equation needs to be solved self-consistently because the self-energy is a function of the solution $\omega$. 
In the present study, however, the $\omega$-dependency in Eq.~(\ref{eq:scph}) can be neglected because we consider only the first-order contribution to the phonon self-energy $\bm{\Sigma}_{\bm{q}}^{(a)}$, which is independent of $\omega$, as will be described in Sec.~\ref{sec:self_energy}.
Nevertheless, the self-consistency is retained in the SCPH approach because the self-energy is a function of phonon frequencies and polarization vectors, which themselves are updated by diagonalizing the matrix $\bm{V}_{\bm{q}}$.

\subsection{Anharmonic self-energy}

\label{sec:self_energy}

Solving the SCPH equation requires a diagrammatic approximation to the phonon self-energy $\bm{\Sigma}_{\bm{q}}(\omega)$.
In this study, we consider anharmonicity up to the fourth order, i.e. $H=H_{0} + U_{3} + U_{4}$, where $U_{n}$ is the $n$th-order contribution to the potential energy surface expressed in terms of the displacement operator $A$.
This can be obtained by substituting 
\begin{equation}
u_{\mu}(\ell\kappa) = (NM_{\kappa})^{-\frac{1}{2}}\sum_{q}\sqrt{\frac{\hbar}{2\omega_{q}}}A_{q}e_{\mu}(\kappa;q)e^{i\bm{q}\cdot\bm{r}(\ell)}
\end{equation}
for Eq.~(\ref{eq:U_Taylor}), where $q$ labels the phonon modes defined as $q = (\bm{q},j)$ and $-q = (-\bm{q},j)$, and $N$ is the number of $\bm{q}$ points. We then obtain the following result:
\begin{align}
U_{n} &= \frac{1}{n!}\bigg(\frac{\hbar}{2}\bigg)^{\frac{n}{2}}\sum_{\{q\}}\Delta (\bm{q}_{1}+\cdots+\bm{q}_{n})
\frac{\Phi(q_{1};\dots;q_{n})}{\sqrt{\omega_{q_{1}}\cdots\omega_{q_{n}}}} \notag \\
& \hspace{20mm} \times A_{q_{1}}\cdots A_{q_{n}}.
\label{eq:Un_in_A}
\end{align}
The function $\Delta (\bm{q})$ becomes 1 if $\bm{q}$ is an integral multiple of the reciprocal vector $\bm{G}$ and is 0 otherwise.
 $\Phi(q_{1};\dots;q_{n})$ is the reciprocal representation of the $n$th-order IFCs defined by
\begin{align}
&\Phi(q_{1};\dots;q_{n}) \notag \\
&= N^{1-\frac{n}{2}}\sum_{\{\kappa,\mu\}}(M_{\kappa_{1}}\cdots M_{\kappa_{n}})^{-\frac{1}{2}} e_{\mu_{1}}(\kappa_{1};q_{1})\cdots e_{\mu_{n}}(\kappa_{n};q_{n}) \notag \\
&  \times \sum_{\ell_{2},\dots,\ell_{n}}\Phi_{\mu_{1}\dots\mu_{n}}(0\kappa_{1};\dots;\ell_{n}\kappa_{n}) e^{i(\bm{q}_{2}\cdot \bm{r}(\ell_{2})+\cdots+\bm{q}_{n}\cdot \bm{r}(\ell_{n}))}.
\label{eq:Phi_recip}
\end{align}

\begin{figure}
\centering
\includegraphics[width=8.0cm, clip]{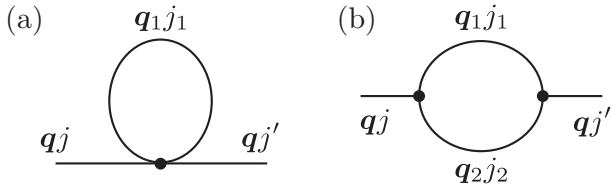}
\caption{Diagrams of the self-energies considered in this study. 
(a) The first-order diagram associated with the quartic term. (b) The second-order diagram associated with the cubic term.}
\label{fig:selfenergy}
\end{figure}

In solving the SCPH equation, we consider only the first-order contribution to the phonon self-energy due to the quartic term
\begin{align}
\Sigma_{\bm{q}jj'}^{(a)}(i\omega_{m}) &= -\frac{1}{2}\sum_{q_{1}}\frac{\hbar\Phi(\bm{q}j;-\bm{q}j';q_{1};-q_{1})}{4\sqrt{\omega_{\bm{q}j}\omega_{\bm{q}j'}}\omega_{q_{1}}} \notag  \\
& \hspace{10mm} \times [1 + 2n(\omega_{q_{1}})], \label{eq:self_a}
\end{align}
which corresponds to the loop diagram shown in Fig.~\ref{fig:selfenergy}(a).
Here, $n(\omega)=[e^{\beta\hbar\omega}-1]^{-1}$ is the Bose-Einstein distribution function.
Since we continue the iteration cycle of the self-consistent equation [Eq.~(\ref{eq:scph})] until we obtain a convergence with respect to the anharmonic frequencies, 
the SCPH equation automatically includes an infinite class of anharmonic self-energies that can be generated from the loop diagram.
In this study, we consider the off-diagonal components of the self-energy to allow for polarization mixing (PM), which we found to be important for c-STO, as will be discussed in Sec.~\ref{sec:scph_STO}.
If we neglect the off-diagonal elements, $\Sigma_{\bm{q}jj'}^{(a)}\approx \Sigma_{\bm{q}jj'}^{(a)}\delta_{jj'}$, the SCPH equation can be reduced to the diagonal form
\begin{align}
\Omega_{q}^{2} &= \omega_{q}^{2} + 2\Omega_{q}I^{(a)}_{q}, \\
I^{(a)}_{q} &= \frac{1}{2}\sum_{q_{1}}\frac{\hbar\Phi(q;-q;q_{1};-q_{1})}{4\Omega_{q}\Omega_{q_{1}}} [1 + 2n(\Omega_{q_{1}})].
\end{align}
This equation is equivalent to the one derived by a variational approach 
where the anharmonic free-energy within the first-cumulant expansion is minimized with respect to trial frequencies~\cite{PhysRevLett.106.165501}.

To calculate the phonon linewidth, one needs to consider the bubble self-energy shown in Fig.~\ref{fig:selfenergy}(b), which is the contribution from cubic anharmonicity given as
\begin{align}
\Sigma^{(b)}_{\bm{q}jj'}(i\omega_{m}) &= \frac{1}{2N}\sum_{q_{1},q_{2}} \frac{\hbar\Phi(-\bm{q}j,q_{1},q_{2})\Phi(\bm{q}j',-q_{1},-q_{2})}{8\sqrt{\omega_{\bm{q}j}\omega_{\bm{q}j'}}\omega_{q_{1}}\omega_{q_{2}}} \notag \\
&\hspace{5mm} \times \Delta(-\bm{q}+\bm{q}_{1}+\bm{q}_{2}) \mathcal{F}(i\omega_{m}, 1, 2).
\label{eq:self_b}
\end{align}
Here we introduced the $\omega$-dependent function $\mathcal{F}$ defined as
\begin{align}
\mathcal{F}(i\omega_{m}, 1,2) &= \sum_{\sigma=\pm 1}\left[ \frac{1+n_{1}+n_{2}}{i\omega_{m}+\sigma(\omega_{1}+\omega_{2})} \right. \notag \\ 
& \left. \hspace{15mm} + \frac{n_{2}-n_{1}}{i\omega_{m}+\sigma(\omega_{1}-\omega_{2})} \right],
\end{align}
where we symbolically denote $n(\omega_{q_{i}})$ and $\omega_{q_{i}}$ as $n_{i}$ and $\omega_{i}$, respectively. 
We will consider the contribution from this diagram in a perturbative manner 
whereby the equation (\ref{eq:self_b}) is evaluated using the phonon frequencies and polarization vectors obtained as a solution to the SCPH equation.
It should be noted that there is another second-order diagram that contains two four-phonon vertexes.
Although we do not consider that contribution for computational reasons, 
it is, in principle, possible to extend the theory to include higher-order diagrams~\cite{Tripathi1974}.

\subsection{Computational implementation}

In this section we describe the details of the computational implementation used to solve the SCPH equation efficiently.
The most expensive part of the SCPH equation is the calculation of the quartic coefficients in Eq.~(\ref{eq:self_a}),
which are changed in each cycle of the iterative algorithm through an update of the phonon eigenvectors.
To avoid recalculating the quartic coefficient in each cycle, we employ a unitary transformation of the eigenvectors, as will be described below.
Our approach is inspired by the method proposed by Hermes and Hirata for molecules~\cite{doi:10.1021/jp4008834},
which we extended to periodic systems at finite temperatures.

First, we construct the dynamical matrix $\bm{D}(\bm{q})$ from the harmonic IFCs and 
calculate eigenvalues and eigenvectors $\{\omega_{q}^{2}, e_{\mu}(\kappa;q)\}$ for the 
gamma-centered $N_{1}\times N_{2}\times N_{3}$ $\bm{q}$-point grid. 
We then calculate the matrix elements $F_{\bm{q}\bm{q}_{1},ijk\ell}= \Phi(\bm{q}i;-\bm{q}j;\bm{q}_{1}k;-\bm{q}_{1}\ell)$ by Eq.~(\ref{eq:Phi_recip}) using the harmonic eigenvectors and quartic IFCs. 
Here, the index $\bm{q}$ is restricted to the irreducible points that are commensurate with the supercell size,
whereas the index $\bm{q}_{1}$ includes all of the $N_{1}\times N_{2}\times N_{3}$ grid points.
The next step is to diagonalize the following SCPH equation, which can be obtained from Eqs.~(\ref{eq:scph}) and (\ref{eq:self_a}):
\begin{equation}
V_{\bm{q}ij}^{[1]} = \omega_{\bm{q}i}^{2}\delta_{ij}+\frac{1}{2}\sum_{\bm{q}_{1},k}F_{\bm{q}\bm{q}_{1},ijkk}
\frac{\hbar\left[1+2n(\omega_{\bm{q}_{1}k})\right]}{2\omega_{\bm{q}_{1}k}}.
\label{eq:scph_first}
\end{equation}
Here we added the superscript 1 to the matrix $\bm{V}_{\bm{q}}$ to explicitly show that it is the first iteration of the SCPH equation.
Then, by diagonalizing the Hermitian matrix $\bm{V}_{\bm{q}}$ as $\bm{V}_{\bm{q}} = \bm{C}_{\bm{q}}\bm{W}_{\bm{q}}\bm{C}_{\bm{q}}^{\dagger}$,
we obtain the updated phonon frequencies $\omega_{\bm{q}i}^{[1]} = W^{\frac{1}{2}}_{\bm{q}ii}$. The corresponding polarization vectors can be obtained from the unitary matrix $\bm{C}_{\bm{q}}$. Let $\bm{E}_{\bm{q}}$ and $\bm{E}_{\bm{q}}^{[1]}$ denote the $s\times s$ matrices defined as $\bm{E}_{\bm{q}}=(\bm{e}_{\bm{q}1},\dots,\bm{e}_{\bm{q}s})$ and $\bm{E}_{\bm{q}}^{[1]}=(\bm{e}_{\bm{q}1}^{[1]},\dots,\bm{e}_{\bm{q}s}^{[1]})$, respectively, where $s$ is the number of phonon modes.
It can then be shown that $\bm{E}_{\bm{q}}$ and $\bm{E}_{\bm{q}}^{[1]}$ are unitary transformations of each other, which can be written as $\bm{E}_{\bm{q}}^{[1]}=\bm{E}_{\bm{q}}\bm{C}_{\bm{q}}^{[1]}$.
Because the phonon polarization vectors are modified in this manner, we need to modify Eq.~(\ref{eq:scph_first}) for the next iteration of the SCPH equation.
The equation for the $n$th step of the iteration is given as
\begin{equation}
V_{\bm{q}ij}^{[n]} = \omega_{\bm{q}i}^{2}\delta_{ij}+\frac{1}{2}\sum_{\bm{q}_{1},k,\ell}F_{\bm{q}\bm{q}_{1},ijk\ell}\mathcal{K}_{\bm{q}_{1},k\ell}^{[n-1]},
\label{eq:V_iter_n}
\end{equation}
where $\mathcal{K}$ is defined as
\begin{align}
\mathcal{K}_{\bm{q},ij}^{[n]} &= \alpha K_{\bm{q},ij}^{[n]} + (1-\alpha) K_{\bm{q},ij}^{[n-1]}, \label{eq:K_mix} \\
K_{\bm{q},ij}^{[n]} &= \sum_{k} C_{\bm{q},ki}^{[n]} C_{\bm{q},kj}^{[n]*} \braket{Q_{\bm{q}k}^{[n]}Q_{\bm{q}k}^{[n]*}} \notag \\ 
&= \sum_{k} C_{\bm{q},ki}^{[n]} C_{\bm{q},kj}^{[n]*} \frac{\hbar\big[1+2n(\omega_{\bm{q}_{1}k}^{[n]})\big]}{2\omega_{\bm{q}_{1}k}^{[n]}}.
\label{eq:K_iter_n}
\end{align}
Here we have used the fact that the mean square displacement of normal coordinate $Q_{\bm{q}j}$ is given as 
$\braket{Q_{\bm{q}j}Q_{\bm{q}j}^{*}} = \frac{\hbar}{2\omega_{\bm{q}j}}\braket{A_{\bm{q}j}A_{\bm{q}j}^{\dagger}}=\frac{\hbar}{2\omega_{\bm{q}j}}[1+2n(\omega_{\bm{q}j})]$. 
In the classical limit $(\beta \rightarrow 0)$, the expectation value would be $\braket{Q_{\bm{q}j}Q_{\bm{q}j}^{*}} = kT\omega_{\bm{q}j}^{-2}$.
In addition, we introduced the mixing parameter $\alpha$ in Eq.~(\ref{eq:K_mix}) to improve convergence.

After we obtain $\bm{V}_{\bm{q}}^{[n]}$ and $\bm{C}_{\bm{q}}^{[n]}$ for all irreducible $\bm{q}$ points, 
we construct the new dynamical matrix as
\begin{align}
\bm{D}_{\bm{q}}^{[n]} &= \bm{E}_{\bm{q}}^{[n]}\bm{W}_{\bm{q}}^{[n]}\bm{E}_{\bm{q}}^{[n]\dagger} \notag \\
&= \bm{E}_{\bm{q}}\bm{C}_{\bm{q}}^{[n]}\bm{W}_{\bm{q}}^{[n]}\bm{C}_{\bm{q}}^{[n]\dagger}\bm{E}_{\bm{q}}^{\dagger},
\label{eq:Dymat_iter_n}
\end{align}
where $W_{\bm{q}ij}^{[n]}=(\omega_{\bm{q}i}^{[n]})^{2}\delta_{ij}$ is the diagonal matrix.
Using the dynamical matrices, we construct dynamical matrices for the star of $\bm{q}$ using the unitary transformation:
\begin{equation}
\bm{D}_{S\bm{q}}^{[n]} = \bm{\Gamma}_{\bm{q}}(\{S|\bm{v}(S)\})\bm{D}_{\bm{q}}^{[n]}\bm{\Gamma}_{\bm{q}}^{\dagger}(\{S|\bm{v}(S)\}).
\end{equation}
Here, $\bm{\Gamma}_{\bm{q}}$ is the unitary matrix associated with the symmetry operation $\{S|\bm{v}(S)\}$ where $S$ is the 3$\times$3 rotation matrix and $\bm{v}(S)$ is the translation vector. The detailed expression for $\bm{\Gamma}_{\bm{q}}$ can be found in Ref.~\onlinecite{RevModPhys.40.1}.
Finally, we construct the dynamical matrix in real space by taking the inverse Fourier transformation 
\begin{equation}
\bm{D}^{[n]}(\bm{r}(\ell)) = \frac{1}{N}\sum_{\bm{q}}\bm{D}_{\bm{q}}^{[n]}e^{-i\bm{q}\cdot\bm{r}(\ell)},
\label{eq:Dymat_IFFT}
\end{equation}
from which we obtain $\omega_{\bm{q}i}^{[n]}$ and $\bm{C}_{\bm{q}}^{[n]}$ for the dense $N_{1}\times N_{2}\times N_{3}$ grid points,
which are necessary for the next iteration of the SCPH equation, by Fourier interpolation.
For polar semiconductors, the non-analytic part of the dynamical matrix is accounted for using the mixed-space approach~\cite{Wang:2010ks}.

We iterate Eqs.~(\ref{eq:V_iter_n})--(\ref{eq:Dymat_IFFT}) until convergence is achieved for all phonon frequencies at the irreducible $\bm{q}$ points.
We initialize the frequency and the unitary matrix as $\omega_{\bm{q}j}^{[0]} = |\omega_{\bm{q}j}|$ and $C_{\bm{q},ij}^{[0]} = \delta_{ij}$, respectively. Whenever we encounter an imaginary branch, we replace the frequency with its absolute value.
After the calculation has converged, the anharmonic frequencies and eigenvectors for a dense $\bm{q}$ grid, which are necessary for the subsequent calculation of phonon lifetime and  lattice thermal conductivity, can be obtained by Fourier interpolation.

\section{Simulation details}
\label{sec:simulation}

\subsection{DFT calculations}
\textit{Ab initio} DFT calculations were performed using the \textit{Vienna ab initio simulation package} (\textsc{vasp})~\cite{Kresse1996}, which employs the projector augmented wave (PAW) method~\cite{PAW1994,Kresse1999}.
The adapted PAW potentials treat the Sr $4s^{2}4p^{6}5s^{2}$, Ti $3s^{2}3p^{6}3d^{2}4s^{2}$, and O $2s^{2}2p^{4}$ shells as valence states. 
A cutoff energy of 550 eV was employed and the Brillouin zone integration was performed with the 12$\times$12$\times$12 Monkhorst-Pack $\bm{k}$-point grid. 
We employed the PBEsol exchange-correlation functional~\cite{PBEsol}, which was reported to work exceedingly well for predicting equilibrium volume and harmonic phonon frequency of BaTiO$_{3}$ and SrTiO$_{3}$~\cite{Kresse_functional}.
The optimized lattice constant is 3.896~{\AA}, which agrees well with the experimental value of 3.905~{\AA} (Ref.~\onlinecite{Okazaki1973545}, 293 K) and the previous DFT result of 3.898~{\AA}~\cite{Kresse_functional}.
The non-analytic part of the dynamical matrix is considered in all of the following calculations.
We calculated the Born effective charges and the dielectric tensor of c-STO using DFPT and obtained values of 
$\epsilon^{\infty}=6.35$, $Z^{*}(\mathrm{Sr})=2.55$, $Z^{*}(\mathrm{Ti})=7.35$, $Z^{*}(\mathrm{O})_{\perp}=-2.04$, 
and $Z^{*}(\mathrm{O})_{\parallel}=-5.82$, which agree well with the previous computational result~\cite{Lasota1997}.  
Because the thermal expansion coefficient of c-STO is very small~\cite{Okazaki1973545},
we neglect thermal expansion effects in this study.

\subsection{Estimation of force constants}
To compute the harmonic phonon frequency, we extracted harmonic IFCs using the finite-displacement approach~\cite{Esfarjani2008}. 
The calculation was conducted with a 2$\times$2$\times$2 cubic supercell containing 40 atoms as in Ref.~\onlinecite{PhysRevB.89.094109}. We displaced an atom from its equilibrium position by 0.01~{\AA} and calculated atomic forces for each displaced configuration. We then extracted $\Phi_{\mu\nu}(\ell\kappa;\ell'\kappa')$ by solving the least-square problem
\begin{equation}
\tilde{\bm{\Phi}} = \arg\min_{\bm{\Phi}} \|A\bm{\Phi}-\bm{F}\|^{2}_{2},
\label{eq:LSE}
\end{equation}
as implemented in the \textsc{alamode} package~\cite{Tadano2014,alamode}. 
Here, $\bm{\Phi}=[\Phi_{1},\Phi_{2},\dots,\Phi_{M}]^{\mathrm{T}}$ is the parameter vector composed of $M$ linearly independent IFCs, $\bm{F}$ is the vector of atomic forces obtained by DFT calculations, and $A$ is the matrix composed of the atomic displacements. 

To solve the SCPH equation and estimate the anharmonic phonon frequencies of c-STO, one has to prepare quartic IFCs. 
Cubic IFCs are also necessary to estimate phonon linewidth and thermal conductivity, as will be discussed in Sec.~\ref{sec:thermal_conductivity}. In principle, one can extend the finite-displacement approach to extract anharmonic terms, for which multiple atoms have to be displaced simultaneously by an appropriately chosen displacement magnitude $\Delta u$. 
However, finding an optimal value of $\Delta u$ is not a trivial task, especially when imaginary modes exist within the harmonic approximation, as in c-STO.
We found that the finite-displacement approach with $\Delta u = 0.1$~{\AA} failed to yield reliable fourth-order IFCs that could reproduce the double-well potential of the AFD mode. 
To avoid this issue, one may alternatively employ the AIMD simulation to sample the displacement-force data set. 
This approach works particularly well for simple systems such as Si and Mg$_{2}$Si~\cite{Tadano2014}. 
However, it should be noted that as long as one employs the ordinary least-squares method [Eq.~(\ref{eq:LSE})], 
an overfitting issue may arise unless the number of individual reference data is fairly large compared with the number of parameters.
Recently, Zhou \textit{et al.}~\cite{PhysRevLett.113.185501} proposed a more robust approach to estimate anharmonic IFCs. 
Noting that only a small fraction of IFCs has non-negligible contributions to atomic forces, 
they developed the compressive sensing lattice dynamics method and obtained the sparse solution 
using the least absolute shrinkage and selection operator (LASSO) technique. 
In the LASSO technique, one solves the following equation:
\begin{equation}
\tilde{\bm{\Phi}} = \arg\min_{\bm{\Phi}} \|A\bm{\Phi}-\bm{F}\|^{2}_{2} + \lambda \|\bm{\Phi}\|_{1},
\label{eq:LASSO}
\end{equation}
where the $L_{1}$ penalty term is added to the least-squares equation. 
Owing to the $L_{1}$ penalty term, one can find a sparse representation of the basis function, 
as demonstrated by the cluster expansion method and the potential fitting~\cite{PhysRevB.87.035125,PhysRevB.90.024101}.
In this work, we followed the procedure of the previous study of Zhou \textit{et al.} to solve the LASSO equation.
We initially conducted an AIMD simulation at 500 K for 2000 steps with the time step of 2 fs.
From the trajectory of the AIMD simulation, we then sampled 40 atomic configurations that were equally spaced in time.
For each configuration, we displaced all of the atoms within the supercell by 0.1~{\AA} in random directions. 
The atomic forces for the configurations prepared in this manner were calculated using precise DFT calculations, from which the matrix $A$ and the vector $\bm{F}$ in Eq.~(\ref{eq:LASSO}) were constructed.
The LASSO equation was solved using the split Bregman algorithm~\cite{PhysRevB.87.035125,doi:10.1137/080725891}, and the optimal value of $\lambda$ was selected from the four-fold cross-validation score.
To ensure that all of the terms in the $L_{1}$ term had the same dimension, we scaled the $n$th-order IFCs and atomic displacement by $\Phi\rightarrow \Phi u_{0}^{n-1}$ and $u\rightarrow u/u_{0}$ respectively, 
with $u_{0} = 0.4~a_{0}$ ($\approx$ 0.21~{\AA}) representing the order of the thermal nuclear motion.

\section{Results and discussion}
\label{sec:results}
\subsection{Anharmonic force constants in cubic SrTiO$_{3}$}

To find a sparse representation of the basis function for c-STO, we first prepared a large parameter vector $\bm{\Phi}$ that included anharmonic terms up to the sixth order. 
For harmonic and cubic terms, we included all possible IFCs present in the 2$\times$2$\times$2 supercell. The quartic terms were considered up to third-nearest neighbor shells, whereas fifth- and sixth-order IFCs were considered for nearest-neighbor pairs. 
We determined a set of linearly independent parameters by considering the space group symmetry and the constraints for the translational invariance~\cite{Tadano2014,Esfarjani2008}.
We fixed the harmonic terms to the values determined by the finite-displacement approach [Eq.~(\ref{eq:LSE})] and employed the LASSO technique for estimating the remaining anharmonic terms. 
The number of linearly independent anharmonic parameters $M$ was 1053, from which a sparse representation was found by Eq.~(\ref{eq:LASSO}).

\begin{figure}
\centering
\includegraphics[width=8.0cm, clip]{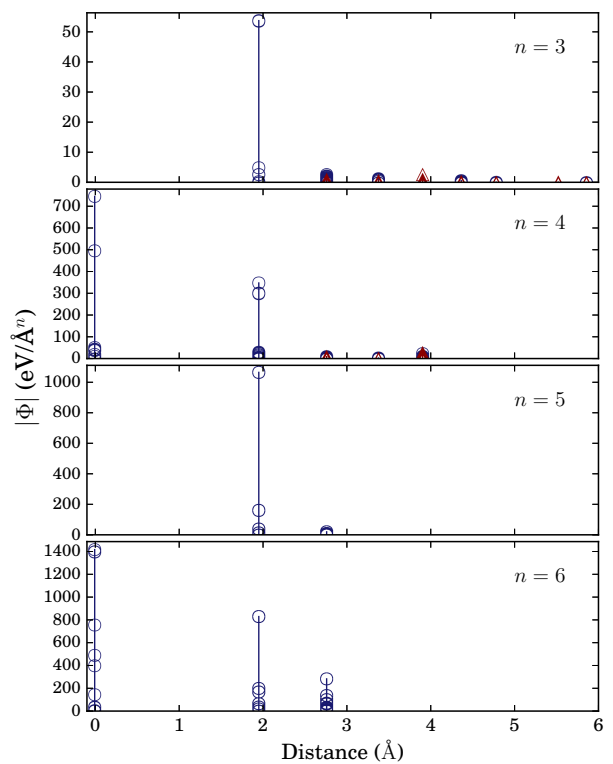}
\caption{(color online). Absolute values of the third-, fourth-, fifth-, and sixth-order anharmonic force constants estimated by the LASSO technique plotted as a function of interatomic distance. The onsite and two-body terms are indicated by circles and the three-body terms are indicated by triangles.}
\label{fig:fcs_lasso}
\end{figure}

\begin{figure}
\centering
\includegraphics[width=8.5cm, clip]{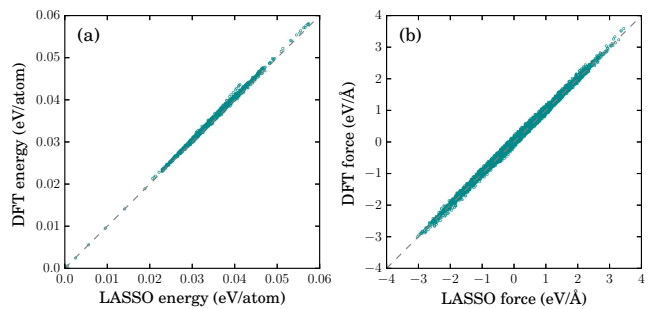}
\caption{(color online). Comparison of (a) potential energy and (b) atomic forces sampled by an individual AIMD simulation at 300 K. 
The dashed lines indicate cases where the results are identical.}
\label{fig:energy_force}
\end{figure}

Figure~\ref{fig:fcs_lasso} shows the magnitude of the anharmonic IFCs estimated by solving the LASSO equation.
Here, the distance for the IFCs related to more than two atoms is defined as the distance of the most distant atomic pairs.
The absence of onsite force constants for the third- and fifth-order IFCs is due to the inversion symmetry of c-STO.
As shown in Fig.~\ref{fig:fcs_lasso}, the magnitude of anharmonic IFCs decays rapidly with increasing interatomic distance, which indicates the locality of anharmonic interactions. 
The terms with the largest magnitude occur at a distance of 1.95~{\AA} and represent force constants between a Ti atom and one of the surrounding O atoms.
Among the onsite quartic terms, $\Phi_{\mathrm{Ti,Ti,Ti,Ti}}^{\mu\mu\mu\mu}$ ($\mu=x,y,z$) and $\Phi_{\mathrm{O,O,O,O}}^{\nu\nu\nu\nu}$, where $\nu$ is the direction parallel to the Ti-O bond, are most significant. 
Compared with these terms, the other onsite IFCs, including those of the Sr atom, are one order of magnitude smaller.

The accuracy of the IFCs estimated by the LASSO equation was assessed by preparing independent test data using an AIMD simulation at 300 K for 2000 steps.
We then calculated the potential energy [Eq.~(\ref{eq:U_Taylor})] and atomic forces using the atomic displacements $\{u\}$ and the IFCs $\{\Phi\}$.
In Fig.~\ref{fig:energy_force} we compare the potential energy $U - U_{0}$ and the atomic forces obtained from DFT and with those calculated from the IFCs estimated by LASSO. 
The model potential well reproduced the DFT results for various atomic configurations.
The relative errors for the test data were 1.4 and 6.1\% for the potential energy and the atomic force, respectively, which are as small as those reported in Ref.~\cite{PhysRevLett.113.185501}.

\subsection{SCPH solution}

\label{sec:scph_STO}

Using the harmonic and quartic force constants obtained from the finite-displacement and the LASSO techniques, respectively, 
the SCPH equation [Eqs.~(\ref{eq:V_iter_n})--(\ref{eq:Dymat_IFFT})] was solved numerically.
Since we employed the 2$\times$2$\times$2 supercell in this study, the $\bm{q}$ point in Eq.~(\ref{eq:V_iter_n}) was limited to the irreducible points on the  2$\times$2$\times$2 grid. 
We changed the $\bm{q}_{1}$ grid to investigate the convergence of the anharmonic phonon frequencies. 
The mixing parameter of $\alpha = 0.1$ was employed for all temperatures except those near the critical temperature of the structural phase transition, where a much smaller $\alpha$ was required.

\begin{figure}
\centering
\includegraphics[width=8.0cm, clip]{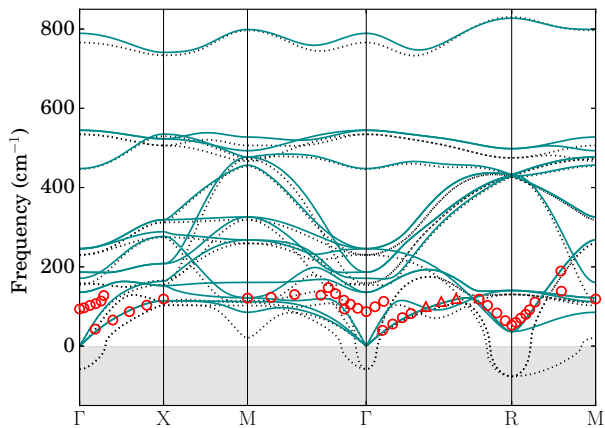}
\caption{(color online). Anharmonic phonon dispersion of c-STO at 300 K calculated using the SCPH theory with 8$\times$8$\times$8 $\bm{q}_{1}$ points (solid lines). The dotted lines show the harmonic phonon dispersion and the open symbols are experimental values at room temperature adapted from Refs.~\onlinecite{Stirling1972,Cowley1969181}.}
\label{fig:scph_band}
\end{figure}

Figure~\ref{fig:scph_band} shows the anharmonic phonon dispersion of c-STO at 300 K obtained as the solution for the SCPH equation. 
The phonon frequencies are increased by the quartic anharmonicity, evident in the low-energy soft modes at the $\Gamma$ $(0,0,0)$, R $(\frac{1}{2},\frac{1}{2}, \frac{1}{2})$, and M $(\frac{1}{2},\frac{1}{2}, 0)$ points.
We investigated the convergence of the anharmonic phonon frequency $\Omega_{q}$ with respect to the 
number of $\bm{q}_{1}$ points. The results for the lowest-energy soft modes at $\Gamma$, R, and M points are summarized in table~\ref{table:anharm_sc1}.
Our results indicate that at least 8$\times$8$\times$8 $\bm{q}_{1}$ points are needed to obtain convergence and a less dense 2$\times$2$\times$2 $\bm{q}_{1}$-point grid significantly overestimates the $\Omega_{q}$ values.
This occurs because the anharmonic phonon-phonon interaction is limited only between the zone-center and zone-boundary phonons by the 2$\times$2$\times$2 $\bm{q}_{1}$ grid. Thus, our numerical results indicate the importance of including mode coupling between longer-wavelength phonons to obtain a reliable description of the phonon softening in c-STO. 
The same size-dependence should also be inherent in the real-space approaches because the available phonon modes are limited by the size of the employed supercell.

\begin{table}[b]
\caption{\label{table:anharm_sc1}  Anharmonic phonon frequency (cm$^{-1}$) of the soft modes at 300 K calculated using the SCPH equation with various $\bm{q}_{1}$-grid densities. The harmonic phonon frequency is also shown for comparison.}
\begin{ruledtabular}
\begin{tabular}{cccc}
 $\bm{q}_{1}$ points & $\Gamma_{15}$ (FE) & R$_{25}$ (AFD) & M$_{3}$ \\
 \hline
 2$\times$2$\times$2 &  144 & 69 & 103 \\
4$\times$4$\times$4 &  138 & 46 &  89 \\
6$\times$6$\times$6 &  136 & 39 &  86 \\
8$\times$8$\times$8 &  136 & 37 &  85 \\
10$\times$10$\times$10 &  135 & 36 &  85 \\
12$\times$12$\times$12 &  135 & 35 &  85 \\
\hline
Frozen phonon & 58$i$ &  76$i$ & 21
\end{tabular}
\end{ruledtabular}
\end{table} 

\begin{figure}
\centering
\includegraphics[width=8.0cm, clip]{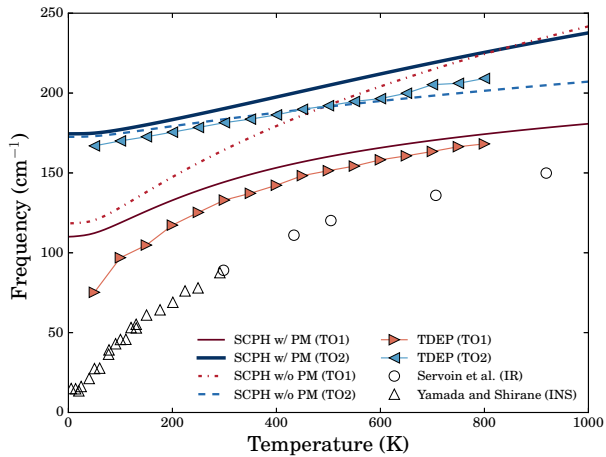}
\caption{(color online). Temperature-dependence of the anharmonic phonon frequencies of two $\Gamma_{15}$ modes calculated using the SCPH equation with and without PM, and with the TDEP approach (see the text for details). The TO1 mode corresponds to the FE mode. 
The open symbols are experimental values for the TO1 mode reported by Servoin \textit{et al.}~\cite{PhysRevB.22.5501} and Yamada and Shirane~\cite{doi:10.1143/JPSJ.26.396}.}
\label{fig:scph_G15}
\end{figure}

We also considered the role of the off-diagonal elements of the phonon self-energy that cause PM.
In Fig.~\ref{fig:scph_G15}, we compare the anharmonic phonon frequencies of two zone-center optical modes, labeled TO1 and TO2, obtained using the SCPH equation with and without PM.
We have shown the SCPH results with 2$\times$2$\times$2 $\bm{q}_{1}$ points, as these results will subsequently be compared with those obtained using an MD-based approach. 
In the SCPH equation without PM, we neglect the off-diagonal elements of the phonon self-energy [Eq.~(\ref{eq:self_a})],
which is obtained by substituting the unitary matrix $\bm{C}_{\bm{q}}$ in Eqs.~(\ref{eq:K_iter_n}) and (\ref{eq:Dymat_iter_n}) with the identity matrix. Therefore, the polarization vectors are fixed to the initial harmonic values.
Fig.~\ref{fig:scph_G15} demonstrates that PM is vital to describe the anti-crossing of the TO1 and TO2 phonon modes, both of which belong to the same irreducible representation $\Gamma_{15}$.
In the case when we neglect PM, an artificial crossing occurs around 500 K 
and the frequencies significantly deviate from those with PM.
Therefore, we conclude that the harmonic polarization vectors should not be employed to predict anharmonic phonon properties of cubic SrTiO$_{3}$ and other perovskite oxides having the same symmetry.

In Fig.~\ref{fig:scph_G15} we compare results obtained with the SCPH method with those obtained with the temperature-dependent effective potential (TDEP) method~\cite{PhysRevB.87.104111}.
In the TDEP method, atomic displacements and forces are sampled by AIMD simulations at a target temperature and are then used to extract effective harmonic force constants by numerical fitting.
In our TDEP simulations, we performed MD simulations using the Taylor expansion potential [Eq.~(\ref{eq:U_Taylor})] instead of AIMD to reduce computational costs.
Anharmonic terms up to the sixth order were considered and the force constants estimated by the LASSO technique were employed. 
We conducted the constant-temperature MD simulations with the 2$\times$2$\times$2 supercell and the temperature was controlled by the Berendsen thermostat~\cite{Berendsen1984}. We employed a time step of 1 fs and conducted the MD simulations for 50000 steps at each temperature. 
The last 40000 steps were employed to extract effective harmonic IFCs by least-squares fitting [Eq.~(\ref{eq:LSE})].
Although the anharmonic frequencies obtained using the TDEP approach are slightly smaller than the SCPH results, 
they agree qualitatively with the SCPH results, as shown in Fig.~\ref{fig:scph_G15}.
We consider this discrepancy to be reasonable for the following two reasons.
First, the SCPH results include only anharmonic self-energies that can be generated from 
Fig.~\ref{fig:selfenergy}(a), whereas the TDEP includes higher-order anharmonic effects. Among these higher-order terms, the first-order contribution due to the cubic anharmonicity, as depicted in Fig.~\ref{fig:selfenergy}(b), should have the largest contribution. 
We found that the effect of the diagram in Fig.~\ref{fig:selfenergy}(b) is to reduce the anharmonic frequency for the FE mode.
Second, since the quantum effect of nuclear motion is not considered in the MD simulation, the thermal average of the squared normal coordinate $\Braket{Q_{q}^{*}Q_{q}}$ is underestimated for temperatures below the Debye temperature in the TDEP approach.
Therefore, the renormalization of anharmonic effects is underestimated in the TDEP approach, which explains why the deviation from the SCPH result becomes larger with decreasing temperature (see TO1 mode in Fig.~\ref{fig:scph_G15}).

\begin{figure}
\centering
\includegraphics[width=8.0cm, clip]{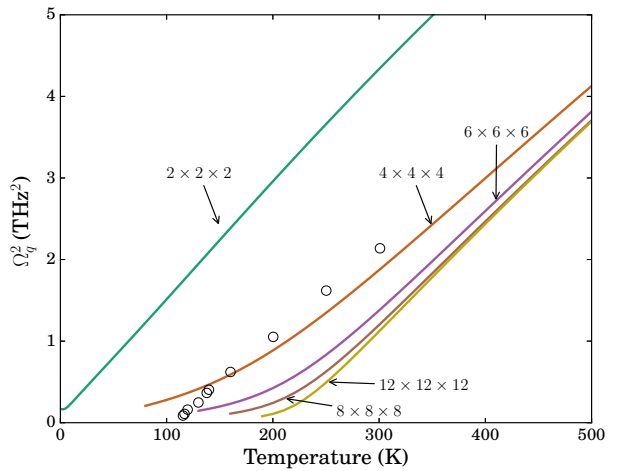}
\caption{(color online). Temperature-dependence of the squared phonon frequency of the R$_{25}$ mode obtained from the SCPH theory with various $\bm{q}_{1}$-point densities. The open circles are experimental values adapted from Ref.~\onlinecite{Cowley1969181}.}
\label{fig:scph_R25}
\end{figure}

Figure~\ref{fig:scph_R25} shows a comparison of the temperature-dependence of the squared frequency of the AFD mode and experimental measurements~\cite{Cowley1969181}.
As can be seen in the figure, the frequency of the AFD mode is severely size-dependent.
For the 2$\times$2$\times$2 $\bm{q}_{1}$ grid, we do not observe a \textit{freezing-out} of the AFD mode even at absolute zero. 
When we increase the $\bm{q}_{1}$-grid density and allow 
interactions with longer-wavelength phonons, we observe the precursor of the freezing-out of the AFD mode at temperatures near 200 K. 
Although the soft-mode frequency does not reach zero in the current simulation with finite $\bm{q}_{1}$ points, 
this would occur in the thermodynamic limit ($N\rightarrow\infty)$ as discussed by Cowley~\cite{Cowley1996585}.
Above approximately 300 K, the temperature-dependence can be reliably fitted by the equation $\Omega_{q}^{2}(T) = a (T-T_{\mathrm{c}})^{2}$.
Applying this equation to the result obtained using the 12$\times$12$\times$12 $\bm{q}_{1}$ grid, we obtain the $T_{\mathrm{c}}$ of the cubic-to-tetragonal phase transition as 220 K.

For comparison, we have plotted experimental results in Figs.~\ref{fig:scph_band},~\ref{fig:scph_G15}, and \ref{fig:scph_R25} using open symbols.
The SCPH equation reproduces the temperature-dependence of the soft modes qualitatively, but not  quantitatively, i.e., the frequencies of the
FE and AFD modes are overestimated and underestimated, respectively.
Because the ADF frequency is underestimated, the transition temperature predicted is twice as large as the experimental value of 105 K.
We consider this deviation to be acceptable because phonon-related properties of ferroelectric materials are known to be sensitive to the lattice constant and exchange-correlation functional employed~\cite{PhysRevB.52.6301,Kresse_functional}.
In this study, we employed the PBEsol functional to avoid problems inherent to the local-density approximation (LDA) and the generalized-gradient approximation with the Perdew-Burke-Ernzerhof parameterization (PBE)~\cite{PhysRevLett.77.3865};
LDA tends to underestimate the equilibrium volume, whereas PBE tends to overestimate it.
However, our numerical results suggest that PBEsol cannot give a quantitative description of c-STO. 
This issue is expected to be resolved, at least partially, by employing a hybrid functional.
Wahl \textit{et al.}~\cite{Kresse_functional} investigated the functional dependence of the harmonic frequency in the FE mode of c-STO
and reported the results of 29$i$ and 74$i$ for the PBEsol semilocal and the Heyd-Scuseria-Ernzerhof (HSE) hybrid functionals~\cite{HSE}, respectively. 
Since the harmonic frequency changes as $\omega_{\mathrm{HSE}}^{2} < \omega_{\mathrm{PBEsol}}^{2} <0$,
we expect that the Fock exchange can increase the depth of the double-well potential, thereby decreasing the anharmonic frequency of the FE mode.
Wahl \textit{et al.} also reported that the energy gain for the AFD phase was smaller in HSE than in the PBEsol functional. 
This indicates that the depth of the double-well potential for the AFD mode can be decreased, and the anharmonic frequency can be increased by using HSE instead of PBEsol.
Therefore, we believe that the quantitative accuracy of the present SCPH results could be improved by employing a hybrid functional, which will be the topic of future work.

In the present SCPH calculations we have not considered effects related to the cubic anharmonicity, such as thermal expansion, 
relaxation of internal coordinates and intrinsic frequency shifts due to the bubble diagram.
However, these effects can, in general, become important in severely anharmonic systems~\cite{Cowley1996585}, 
and should be considered, especially when one intends to quantitatively compare theoretical results with experimental data.
Therefore, extending the present \textit{ab initio} method to include these effects, either perturbatively or self-consistently,
could be another important direction for further research and development.

\subsection{Lattice thermal conductivity}
\label{sec:thermal_conductivity}

The lattice thermal conductivity is a key quantity for optimizing the thermoelectric figure-of-merit $ZT$, and it has been the subject of intense theoretical study in recent years.
To show the validity of our theoretical approach based on the SCPH equation, 
we estimated the lattice thermal conductivity of c-STO.
For this work, we employ the Boltzmann transport equation (BTE) within the relaxation time approximation (RTA), where the lattice thermal conductivity is given as
\begin{equation}
\kappa_{L}^{\mu\nu}(T) = \frac{1}{VN}\sum_{q} C_{q}(T)v_{q}^{\mu}(T)v_{q}^{\nu}(T)\tau_{q}(T).
\label{eq:kappa}
\end{equation}
Here, $V$ is the unit-cell volume, $C_{q}$ is the lattice specific heat, $\bm{v}_{q}=d\Omega_{q}/d\bm{q}$ is the group velocity, and $\tau_{q} = [2\Gamma_{q}(\Omega_{q})]^{-1}$ is the lifetime of phonon $q$.
The phonon linewidth $\Gamma_{q}(\omega)$ can be obtained from the imaginary part of the phonon self-energy that results from the cubic anharmonicity [Eq.~(\ref{eq:self_b})], which is given explicitly as
\begin{align}
\Gamma_{q}(\omega) &= \frac{\pi}{2N}\sum_{q',q''} \frac{\hbar|\Phi(-q,q',q'')|^{2}}{8\Omega_{q}\Omega_{q'}\Omega_{q''}} \notag \\
&\hspace{5mm} \times \left[ (n_{q'} + n_{q''} + 1)\delta(\omega - \Omega_{q'} - \Omega_{q''}) \notag  \right. \\
&\hspace{11mm} \left. - 2(n_{q'} - n_{q''}) \delta(\omega - \Omega_{q'} + \Omega_{q''}) \right] .
\label{eq:Gamma}
\end{align}
Here, the matrix element $\Phi(q,q',q'')$ is calculated from cubic IFCs using Eq.~(\ref{eq:Phi_recip}) with eigenvectors $\{e_{\mu}(\kappa;q)\}$  replaced by the solution to the SCPH equation $\{\epsilon_{\mu}(\kappa;q)\}$.
The equations (\ref{eq:kappa}) and (\ref{eq:Gamma}) are identical to those that have commonly been employed in the thermal conductivity calculations 
except that harmonic phonon frequencies and eigenvectors are substituted by anharmonic frequencies and eigenvectors, respectively, obtained using the SCPH equation.

\begin{figure}
\centering
\includegraphics[width=8.0cm, clip]{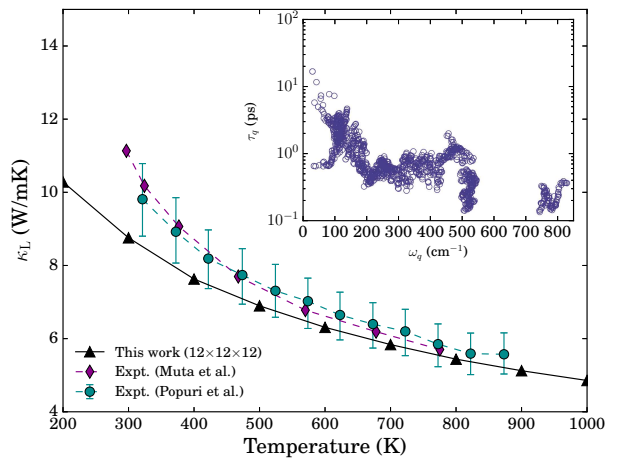}
\caption{(color online). Temperature-dependence of the lattice thermal conductivity of c-STO. 
The computational result is compared with experimental values reported by Muta \textit{et al.}~\cite{Muta2005306} and Popuri \textit{et al.}~\cite{C4RA06871H}. 
Lines are shown to guide the eye. Inset: Calculated phonon lifetime of c-STO at 300 K.}
\label{fig:kappa_and_tau}
\end{figure}

Figure~\ref{fig:kappa_and_tau} compares the calculated thermal conductivity of c-STO with experimental results~\cite{Muta2005306,C4RA06871H}.
The calculation was conducted using the 8$\times$8$\times$8 $\bm{q}_{1}$ grid for the SCPH equation and
the 12$\times$12$\times$12 $\bm{q}$ grid for the BTE-RTA equation [Eq.~(\ref{eq:kappa})].
Although we observed deviations in soft-mode frequencies, 
the calculated thermal conductivity agrees well with the experimental results, as can be seen in Fig.~\ref{fig:kappa_and_tau}.
We expect that the agreement could be improved further by employing a finer $\bm{q}$ grid and using a hybrid functional, although such calculations were not performed because of computational limitations.
In the Fig.~\ref{fig:kappa_and_tau} inset, we also show the phonon lifetime $\tau_{q}$ at 300 K calculated by Eq.~(\ref{eq:Gamma}).
The phonon lifetimes of c-STO obtained from the perturbation theory [Eq.~(\ref{eq:Gamma})] are found to be even smaller than those of PbTe~\cite{Tian2012}, but the 
$\kappa_{\mathrm{L}}$ value of c-STO is higher.
This can be attributed to the large group velocities of phonons, especially of TO modes above $\sim$ 100 cm$^{-1}$, which contribute significantly to the total $\kappa_{L}$ value~\cite{Feng2015}.
The lifetime shows a characteristic feature in the low-frequency region ($<$ 100 cm$^{-1}$): the phonon modes split into two separate regions in $\tau_{q} > 3$ ps and $\tau_{q} \sim 0.6$ ps, where the former corresponds to the acoustic modes that follow the frequency dependence of $\tau\sim\omega^{2}$, 
which has been observed in other materials~\cite{Tadano2014,Tian2012}, and the latter corresponds to the phonon modes around the R point, 
which indicates the severe anharmonicity of the AFD mode.

\section{Conclusions}
\label{sec:conclusion}

We developed an \textit{ab initio} method to compute anharmonic phonon frequencies and lifetimes that can be applied to severely anharmonic systems.
The method employs anharmonic force constants up to the fourth order,
which are extracted from DFT calculations using a compressive sensing approach.
The frequency renormalization associated with the quartic anharmonicity 
is treated non-perturbatively using the SCPH theory.
By performing the perturbation calculation after the SCPH solution, 
we also calculated phonon lifetimes that result from the three-phonon scattering processes.

We applied the method to the high-temperature phase of perovskite SrTiO$_{3}$.
Unlike the harmonic phonon dispersion, the SCPH solution was free from the imaginary branches in the entire Brillouin zone.
We found that including polarization mixing is important to correctly account for the temperature dependence of the phonon frequency of the ferroelectric soft mode of perovskite oxides.
In addition, we examined the size-dependence of the anharmonic frequencies of the soft modes and found that long-wavelength phonons significantly reduced the anharmonic frequencies, especially for the antiferrodistortive mode near the transition temperature. 
The temperature-dependence of the soft mode frequencies calculated using the SCPH theory agreed 
qualitatively well with the experimental results.
However, the quantitative accuracy of the present calculations based on the PBEsol functional was unsatisfactory, 
where we obtained the cubic-to-tetragonal transition temperature as $T_{\mathrm{c}} = 220$ K that was twice as large as the experimental value of 105 K. 
Although further theoretical investigations are required to understand the origin of this discrepancy, 
we expect that the quantitative agreement can be improved by employing a hybrid functional.
We also calculated the lattice thermal conductivity $\kappa_{\mathrm{L}}$ of cubic SrTiO$_{3}$ using the Boltzmann transport equation within the relaxation-time approximation.
The calculated $\kappa_{\mathrm{L}}$ values reproduced experimental results especially in the high temperature region.
The underestimation of $\kappa_{\mathrm{L}}$ in the low temperature region may be attributed to 
the overestimation (underestimation) of the ferroelectric (antiferrodistortive) soft mode, 
which will be addressed in a future work.

The present method, which combines the SCPH theory with perturbation approach based on anharmonic force constants, 
enables us to obtain the anharmonic phonon frequencies and lifetimes at various temperatures efficiently just by changing the occupation number. 
The system size dependency can be investigated using the reciprocal space formalism.
Therefore, we believe that the present method paves the way for understanding lattice anharmonicity and related dynamical and thermodynamical properties of thermoelectric, ferroelectric, and superconducting materials.

\section*{Acknowledgements}

We wish to thank Takashi Miyake, Mitsuaki Kawamura, and Takuma Shiga for fruitful discussions, and Masato Okada for helpful suggestions regarding the compressive sensing. 
This study is partially supported by Tokodai Institute for Element Strategy (TIES) funded by MEXT Elements Strategy Initiative to Form Core Research Center and also by Thermal Management Materials and Technology Research Association (TherMAT).
The computation in this work has been done using the facilities of the Supercomputer Center, Institute for Solid State Physics, The University of Tokyo.


%

\end{document}